\documentstyle[12pt]{article}
\def\thebibliography#1{\leftline{\it References}\list
  {[\arabic{enumi}]}{\settowidth\labelwidth{[#1]}\leftmargin\labelwidth
    \advance\leftmargin\labelsep
    \usecounter{enumi}}
    \def\newblock{\hskip .11em plus .33em minus .07em}
    \sloppy\clubpenalty4000\widowpenalty4000}

\newcommand{\be}{\begin{eqnarray}}
\newcommand{\ee}{\end{eqnarray}}
\newcommand{\dslash}{\partial \hskip -0.5em /}
\newcommand{\Dslash}{D \hskip -0.7em /}
\newcommand{\Vslash}{V \hskip -0.7em /}

\newcommand{\Aslash}{A \hskip -0.7em /}
\newcommand{\tr}{{\rm tr}}
\newcommand{\Tr}{{\rm Tr}}

\newcommand{\La}{{\cal L}}
\newcommand{\A}{{\cal A}}

\newcommand{\ie}{{\it i.e.}\ }
\newcommand{\eg}{{\it e.g.}\ }
\newcommand{\cf}{{\it cf.}\ }
\newcommand{\zr}[1]{\mbox{\hspace*{#1em}}}
\newcommand{\ID}{\mbox{{\sf 1}\zr{-0.16}\rule{0.04em}{1.55ex}\zr{0.1}}}
\newcommand{\textlineskip}{\baselineskip=14pt}

\begin{document}

\rightline{UNIT\"U-THEP-14/1994}
\rightline{July 1994}
\rightline{hep-ph/9407305}
\vskip .5truecm
\vskip 2truecm
\centerline{\Large\bf The Skyrmion limit of}
\vskip 0.5truecm
\centerline{\Large\bf the Nambu--Jona-Lasinio soliton}
\baselineskip=20 true pt
\vskip 2cm
\centerline{U.\ Z\"uckert, R.\ Alkofer, H.\ Reinhardt and H.\ Weigel$^\dagger$}
\vskip .3cm
\centerline{Institute for Theoretical Physics}
\centerline{T\"ubingen University}
\centerline{Auf der Morgenstelle 14}
\centerline{D-72076 T\"ubingen, Germany}
\vskip 5.95cm
\centerline{\bf ABSTRACT}
\vskip .25cm

The special role of the isoscalar mesons ($\sigma$ and $\omega$) in
the NJL soliton is discussed. Stable soliton solutions are obtained
when the most general ansatz compatible with vanishing grand spin is
assumed. These solutions are compared to soliton
solutions of a purely pseudoscalar Skyrme type model which is related to
the NJL model by a gradient expansion and the limit of infinitely
heavy (axial-) vector mesons.

\vfill
\noindent
$^\dagger $
{\footnotesize{Supported by a Habilitanden--scholarship of the
Deutsche Forschungsgemeinschaft (DFG).}}
\eject

\normalsize\textlineskip
\leftline{\it Introduction}

The theory of strong interactions, Quantum
Chromo Dynamics (QCD), reduces
for a large number of colors $N_c$ to an effective theory of weakly
interacting meson fields\cite{tHo74}. Furthermore baryons emerge as
soliton solutions in this effective meson theory\cite{Wi79}. On the
other hand, simplified models with quark degrees of freedom can also
possess solitonic solutions, as for example the
Nambu--Jona-Lasinio (NJL) model\cite{Na61}. Selfconsistent soliton
solutions have been found\cite{Re88a}-\cite{Al90} where the
calculations had been restricted
to the chiral field. It has then been observed that a straightforward
extension of this model to allow for space dependent scalar fields
leads to a collapse of the soliton\cite{Wa92,Si92}. Simultaneously the
NJL model containing the chiral as well as (axial-) vector meson
fields was shown to possess stable soliton
solutions\cite{Al91}-\cite{Zu94}. In case the scalar degrees of
freedom are incorporated in the latter type of model, the repulsive
character of the $\omega$ meson is supposed to prevent the soliton
from collapsing.

The aim of the present investigations is twofold.
First, we will discuss the special role of the isoscalar meson fields
related to the above mentioned collapse. With all
meson fields included in the soliton calculations the valence quarks
are strongly bound and they
indeed join the Dirac sea. Therefore the topological current, which
arises in leading order in the gradient expansion of the vacuum part
of the baryon current, becomes a
suitable representation of the full baryon current. This fact supports
the Skyrmion picture of baryons. For these investigations we are going
to employ a static energy functional which
has recently been motivated by studying the relevant analytic
properties under Wick rotation\cite{We94}. For the inclusion of all
mesons similar calculations have previously been performed using a
somewhat different definition of the energy functional\cite{Ru94}. As
the Minkowski energy functional is ambiguous if the $\omega$ meson is treated
non-perturbatively\cite{We94} it is, of course, interesting to study the
question of stability assuming an alternative definition for the
energy functional. Combining the results of the two approaches should
provide a general picture of the situation.

On the other hand it is known that a Skyrme type model can be derived
from the bosonized NJL model\cite{Re89}. This suggests to compare the
self-consistent soliton solution of the NJL model to the soliton
solution of the corresponding Skyrme model. Such a comparison
constitutes the second issue of this letter.

\vskip0.3cm

\leftline{\it Description of the model}

As starting point we assume the bosonized version of the two-flavor NJL
model\cite{Eb86}:
\be
\A &=& \A _F + \A_m,
\\*
\A _F &=& \Tr \log (i\Dslash )
 = \Tr \log \left(i\dslash +\Vslash +\gamma_5\Aslash
- (P_R\Sigma+P_L\Sigma^{\dag}) \right),
\\*
\A _m &=& \int d^4x \left(-\frac{1}{4g_1}
\tr(\Sigma^{\dag}\Sigma-\hat m_0(\Sigma+\Sigma^{\dag})+\hat m_0^2)
+\frac{1}{4g_2}\tr(V_\mu V^\mu+A_\mu A^\mu) \right) .
\ee
Here $V_\mu=\sum_{a=0}^3V_\mu^a\tau^a/2$ and
$A_\mu=\sum_{a=0}^3A_\mu^a\tau^a/2$ denote the vector and axial
vector meson fields. The matrices $\tau ^a/2$ denote the generators of
the flavor group ($\tau^0=\ID$). $P_{R,L} = (1\pm \gamma _5)/2$ are
the projectors on right-- and left--handed quark fields, respectively.
The complex field $\Sigma$ describes the scalar
and pseudoscalar meson fields $S_{ij}=\sum_{a=0}^3S^a\tau^a_{ij}/2$
and $P_{ij}=\sum_{a=0}^3P^a\tau^a_{ij} /2$:
\be
 \Sigma=S + i P = \Phi\ U
\ee
where we have made use of a polar decomposition. This defines the
chiral radius $\Phi$ as well as the chiral field $U=\exp(i \Theta)$
with $\Theta$ being the chiral angle.  The current quark mass
matrix $\hat m_0={\rm diag}(m_0^u,m_0^d)$ only appears in the mesonic
part of the action, ${\cal A}_m$. The Schwinger--Dyson (or gap) equation
relates $\hat m_0$ to the constituent quark mass matrix
$\hat m={\rm diag}(m^u,m^d)$. For the ongoing discussion we will
adopt the isospin limit $m_0^u=m_0^d=:m_0$ which also implies
$m^u=m^d=:m$.  As the NJL model is not renormalizable it needs
regularization. This introduces one more parameter, the cut--off
$\Lambda$. Hence the model contains four parameters: $m_0,g_1,g_2$ and
$\Lambda$. As ingredients from the meson sector we use the pion decay
constant $f_\pi=93$MeV and the masses of the pion and the
$\rho$--meson, $m_\pi=135$MeV and $m_\rho=770$MeV, respectively.
Then the Schwinger--Dyson equation allows one to choose the
constituent quark mass $m$ as the only free parameter and express
$m_0,g_1,g_2$ and $\Lambda$ in terms of it.
We employ Schwinger's proper time description\cite{Sch51} which has
the desired feature of preserving gauge symmetry. Then the fermion
determinant, ${\cal A}_F$, may be decomposed into real (${\cal A}_R$) and
imaginary (${\cal A}_I$) parts. Only ${\cal A}_R$ is divergent. The
proper time procedure is applicable to ${\cal A}_R$ since
$\Dslash_E^{~\dagger}\Dslash_E$ is positive definite
\be
{\cal A}_R=\frac{1}{2}{\rm Tr}{\rm log}\left(
\Dslash_E^{~\dagger}\Dslash_E\right)\rightarrow
-\frac{1}{2}\int_{1/\Lambda^2}^\infty\frac{ds}{s}
{\rm exp}\left(-s\Dslash_E^{~\dagger}\Dslash_E\right).
\label{arai}
\ee
Although the imaginary part is UV finite, the proper time regularization
may be applied as well\cite{Al92a,Zu94}. Note,
however, that regularizing ${\cal A}_I$ or not yields qualitatively
different models and different results may occur.

The definition of the Euclidean Dirac one-particle Hamiltonian $h$ via
\be
i\beta\Dslash_E=-\partial_\tau-h
\label{diracham}
\ee
is useful in the context of studying soliton solutions since
one has $[\partial_\tau,h]=0$ for static fields. In the
limit of large Euclidean times, $T\rightarrow\infty$, the temporal part
of the trace may be carried out straightforwardly by calculating
Gaussian integrals. To evaluate the
trace over the remaining degrees of freedom we require the eigenvalues
of $h$. As $h$ is non--Hermitean, we have to distinguish between left
and right eigenstates
\be
h|\Psi_\nu\rangle = \epsilon_\nu |\Psi_\nu\rangle\quad
\langle\tilde{\Psi}_\nu|h=\epsilon_\nu\langle\tilde{\Psi}_\nu|
\qquad \ie h^\dagger |\tilde{\Psi}_\nu\rangle = \epsilon_\nu^*
|\tilde{\Psi}_\nu\rangle
\label{eigen}
\ee
with the normalization condition
$\langle\tilde{\Psi}_\mu|\Psi_\nu\rangle =
\delta_{\mu\nu}$. The functional trace may then be expressed as sums
over the one--particle energies $\epsilon_\mu$.
The resulting expression motivates the following definition for the Minkowski
energy functional\cite{We94}:
\be
E[\varphi]=E_{\rm val}^R+E_{\rm val}^I+E_{\rm vac}^R+E_{\rm vac}^I
+E_m-E_{\rm vac}^R[\varphi_{\rm vac}]
\label{efunct}
\ee
where for simplicity we have generically labeled the meson fields by
$\varphi$.  The contribution of the trivial vacuum has been subtracted
whereby the configuration $\varphi_{\rm vac}$ corresponds to $\Sigma=m$
while the (axial-) vector meson fields are set to zero.
$E[\varphi]$ receives
contributions from the explicit occupation of the valence quark
level\cite{Re89a,Al94}
\be
E_{\rm val}^R=N_C\sum_\nu\eta_\nu |\epsilon_\nu^R|,
\qquad E_{\rm val}^I=N_C\sum_\nu\eta_\nu
{\rm sgn}(\epsilon_\nu^R)\epsilon_\nu^I,
\label{eval}
\ee
the polarized Dirac sea
\be
E_{\rm vac}^R=\frac{N_C}{4\sqrt{\pi}}\sum_\nu |\epsilon_\nu^R|
\Gamma\big(-\frac{1}{2},(\epsilon_\nu^R/\Lambda)^2\big),\quad
E_{\rm vac}^I=\frac{-N_C}{2}\sum_\nu \epsilon_\nu^I
{\rm sgn} (\epsilon_\nu^R)
\label{evac}
\ee
and the mesonic part, $E_{\rm m}$, which stems from $\A_{\rm m}$.
In eqn.~(\ref{eval}) $\eta_\nu$ denote the occupation
numbers of the valence quark orbits. They are
adjusted to provide unit baryon number:
$1=\sum_\mu\left(\eta_\mu-\frac{1}{2}{\rm sgn}(\epsilon_\mu^R)\right)$.
In eqn.~(\ref{evac}) we have displayed the vacuum part of the energy for the
case that the imaginary part is not regularized; employing the regularization
for the imaginary part yields\cite{Al92a,Zu94}
\be
E_{\rm vac}^I=\frac{-N_C}{2\sqrt{\pi}}\sum_\nu \epsilon_\nu^I
{\rm sgn} (\epsilon_\nu^R)
\Gamma\big(\frac{1}{2},(\epsilon_\nu^R/\Lambda)^2\big).
\label{evacimag}
\ee

It is important to note that the different treatments of real and
imaginary parts under regularization destroys the analytical structure
of the action in the time components of the (axial--) vector meson
fields\footnote{This is manifest in eqns.~(\ref{eval},\ref{evac}) by the
explicit appearance of real and imaginary parts of the energy eigenvalues
$\epsilon_\mu$.}.
As already indicated this prohibits a unique extraction of a Minkowski energy
functional. In ref.~\cite{We94}, however, it has been demonstrated that the
definition (\ref{efunct}) is motivated from the regularized energy
functional in Euclidean space. This motivation is based on the fact
that the energy eigenvalue, $\epsilon_\mu$, may well be approximated
by a sum consisting of two functionals of the meson fields. One of
these two functionals is almost independent of $\omega$ while the
other has only a linear dependence on $\omega$, see tables 2.1 and 2.2
of ref.~\cite{We94}. Furthermore, the energy functional (\ref{efunct})
possesses the correct behavior under global flavor singlet
transformations and it also yields the current field identities. A
unique Minkowski energy functional can only be obtained when expanding
the fermion determinant in terms of the time components of the (axial--)
vector meson fields while treating all other fields as non--perturbative
background fields ({\it cf.} ref.~\cite{We94}).

Let us next construct the most general static Euclidean Dirac Hamiltonian
in the grand spin zero sector. The grand spin operator is defined as the
sum $\mbox{\boldmath $G$}=\mbox{\boldmath $l$}+\mbox{\boldmath
$\sigma$}/2 +\mbox{\boldmath $\tau$}/2$ with $\mbox{\boldmath $l$}$
being the orbital angular momentum, $\mbox{\boldmath $\sigma$}/2$ the
spin and $\mbox{\boldmath $\tau$}/2$ the isospin operators. For the
chiral field the well--known hedgehog {\it ansatz}
\be
U(\mbox {\boldmath $r $})={\rm exp}\Big(i{\mbox{\boldmath $\tau$}}
\cdot{\hat{\mbox {\boldmath $r $}} }\Theta(r)\Big)
\label{ansatz1}
\ee
satisfies the condition of vanishing grand spin while for the
scalar field this can only be accommodated by a radial function
\be
\Phi(\mbox{\boldmath $r$}) = m \phi(r).
\label{ansatz2}
\ee
For the (axial--) vector meson fields we impose the grand spin symmetric
{\it ans\"atze}:
\be
\begin{array}{rllrl}
V_\mu^0&=&\omega(r)\delta_{\mu4},&
\quad V_4 ^a&=0, \quad V_i^a
= \epsilon ^{aki} \hat r^k G(r), \\
A_\mu^0&=&0,&
\quad A_4^a&=0, \quad
A_i^a = \hat r^i \hat r^a F(r) + \delta ^{ia} H(r),
\end{array}
\label{ansatz3}
\ee
where the indices $a,i$ and $k$ take the values 1, 2 and 3.
Then the Euclidean Dirac Hamiltonian reads
\be
h &=& \mbox {\boldmath $\alpha \cdot p $}+i\omega(r)
+ m \phi(r) \beta({\rm cos}\Theta(r)+i\gamma_5
{\mbox{\boldmath $\tau$}}\cdot\hat{\mbox{\boldmath $r$}}{\rm sin}\Theta(r))
\nonumber \\*
&&+\frac 1 2 (\mbox {\boldmath $\alpha $} \times \hat{\mbox{\boldmath $r$}} )
{\mbox{\boldmath $\cdot\tau$}} G(r)
+ \frac 1 2 (\mbox {\boldmath $\sigma \cdot $} \hat{\mbox{\boldmath $r$}} )
(\mbox {\boldmath $\tau  \cdot $} \hat{\mbox{\boldmath $r$}} ) F(r)
+\frac 1 2 (\mbox {\boldmath $\sigma \cdot \tau $} ) H(r).
\label{hamil}
\ee
With the {\it ans\"atze} (\ref{ansatz1}) - (\ref{ansatz3}) the mesonic part of
the
energy is given by
\be
E_{\rm m}&=&4\pi \int dr r^2\Bigl( \frac {m_\pi^2f_\pi^2}{2m_0}
 \bigl[m(\phi^2(r)-1) + 2m_0\phi(r)(1- \cos\Theta(r))\bigr]
\nonumber \\*
&& \qquad
+\big(\frac{m_\rho}{g_V}\big)^2\bigl[G^2(r)+\frac{1}{2}F^2(r)+F(r)H(r)+
\frac{3}{2}H^2(r)-2 \omega^2(r)\bigr]\Bigr).
\label{emes}
\ee
Here $g_V$ is the universal vector coupling constant,
$g_V = \left( \frac 1 {8\pi ^2} \Gamma (0,\frac {m^2}{\Lambda ^2})
\right) ^{-1/2}$ which is related to the coupling constant $g_2$ via
the $\rho$--meson mass $g_V^2=4 g_2 m_\rho^2$\cite{Eb86}.

The equations of motion for the meson fields are derived by extremizing
the static energy functional (\ref{efunct}):
$\delta E[\varphi]/\delta\varphi=0$. In addition to the equations of
motion listed in ref.~\cite{Zu94} we obtain for the scalar field
\be
 \phi(r)=\frac{m_0}{m}\cos{\Theta(r)} -
 \frac{m_0 N_c}{m_\pi^2f_\pi^2} \tr\int \frac {d\Omega}{4\pi}\,
 \Bigl( \cos\Theta(r) + i\gamma_5 {\mbox{\boldmath $\tau$}}\cdot
 \hat{\mbox{\boldmath $r$}} \sin\Theta(r) \Bigr)\rho (\mbox{\boldmath $r,r$}).
\label{eqmscalar}
\ee
with the scalar density matrix $\rho(\mbox{\boldmath $x,y$})$ (which is
bilinear in the eigenfunctions of $h$) being defined in
ref.~\cite{Zu94}. The formal structure of the other equations
of motion is not effected by the presence of the scalar field $\phi$.

In order to diagonalize $h$ and to solve the equations of motion
$\delta E[\varphi]/\delta\varphi=0$ we discretize the Hamiltonian
(\ref{hamil}) in a suitable basis using a spherical cavity of finite
radius $D$. Typical values are $D=4 \ldots 6 \rm{fm}$. For details about the
numerical method see ref.~\cite{Zu94a}. The numerical solution to
$\delta E/\delta\varphi=0$ on the whole range $0 \le r \le D$ is
plagued by finite size effects\cite{Zu94}. In order to avoid these
effects we substitute the exact solution to $\delta E/\delta\varphi=0$
by the phenomenologically motivated large distance behavior of the
meson fields for $r \ge r_m$ ($D/4 \le r_m \le D/2$). For the chiral
angle $\Theta$ this can be extracted from the solution of the free
(P-wave) Klein-Gordon equation:
\be
 \Theta(r) \stackrel{r \to \infty}{\longrightarrow} \tilde{\Theta}(r)
 \sim \frac {e^{-m_\pi r}} {r} \left(m_\pi + \frac 1 r \right).
\label{tail1}
\ee
For the other fields we make use of their relation to the chiral angle in a
local approximation yielding the asymptotic behavior\footnote{In order to
accommodate the vacuum values of the fields at $r=D$ we furthermore multiply
appropriate factors to the RHS of eqns.~(\ref{tail1}) and (\ref{tail2}):
$\varphi \to \varphi \tanh (a (1-\frac r D)) \quad \hbox{with}
 \quad a \approx 10 - 15.$}
\be
 G(r) &\sim& \tilde{\Theta}^2(r), \nonumber \\*
 \omega(r) &\sim& \tilde{\Theta}^\prime(r)\tilde{\Theta}^2(r), \nonumber \\*
 \phi(r)-1 &\sim& \tilde{\Theta}^2(r) + a e^{-2mr}  , \nonumber \\*
 F(r),H(r) &\sim& \tilde{\Theta}^\prime(r).
\label{tail2}
\ee
The constants of proportionality, which are omitted in
eqns.~(\ref{tail1},\ref{tail2}), are fixed by making contact with the exact
solution to $\delta E/\delta\varphi=0$ at $r=r_m$. For the scalar
field we fix the additional constant $a$ by requiring the derivative of
$\phi(r)$ to be continuous.
\vskip0.3cm

\leftline{\it The role of the isoscalar meson fields}
In some recent studies of the NJL model \cite{Wa92,Si92} it has been
shown that abandoning the chiral circle condition $S^2+P^2=m^2$ leads to
the collapse of the soliton configuration if no additional mechanism
to preserve the stability is incorporated (\eg including a forth-order
term in the scalar meson field \cite{We93,Me92} or constraining the
regularized baryon number \cite{Sch93}). It has already been
demonstrated that in an alternative definition of the (ambiguous)
non-perturbative energy functional the incorporation of the
$\omega$ meson provides a further stabilisation mechanism\cite{Ru94}.
This is intuitively clear because the $\omega$ meson is of repulsive
character.

As already remarked in the preceding section regularizing the imaginary part
of the action or not leads to quite different models. Taking a finite
cut-off for the imaginary part (\ref{evacimag}) also yields a
regularized baryon number. As
a matter of fact this regularized baryon number density vanishes for the field
configuration associated with the above mentioned
collapse\cite{Sch93} indicating that the collapse is due to the
transition from the baryon number $N_B=1$ sector to the $N_B=0$
sector. Since the baryon
number density represents the source of the $\omega$ field nothing
prevents the $\omega$ field from being zero when the imaginary part of
the action is regularized and the scalar field is allowed to be space
dependent. Consequently, the collapse also appears if all vector
mesons are included and $\A_I$ is regularized.
The ongoing explorations will therefore be constrained to the case
when $\A_I$ is \underline{not} regularized. Then a non-vanishing source for the
$\omega$ field is present and the stability of the soliton depends
on the strength of the $\omega$ field. It has been noted that
the numerical calculations will pretend to a ``pseudo-stable solution"
if too small a basis for diagonalizing $h$ in momentum space is
adopted\cite{Si92}. By varying the size of this basis we have ensured that our
solutions are not subject to this ``pseudo-stability". At low constituent quark
masses, $m$, the trivial minimum with $N_c$ free and unbound valence
quark orbits occupied is energetically favored compared to the
soliton configuration. Actually a lower bound exists for $m$
below which the equation of motion are solved by this trivial
configuration only. Numerically we obtain for this bound approximately
$330$MeV which it is somewhat larger ($\sim 410$MeV) when neither
$\rho$ nor $a_1$ fields are present. It should be stressed that this
instability is completely different in character from the above
mentioned collapse. In case of the collapse the valence quark is
strongly bound and its energy eigenvalue tends to $-m$.

For stable solitons the valence quark is strongly bound
($\epsilon_{\rm val}^R {_{\displaystyle<} \atop^{\displaystyle\sim}}
-m/4$) and the chiral radius $\phi$ lies near the chiral circle $\phi(r)=1$
(\cf fig.~1). Then the scalar-isoscalar meson field has
only minor influence on the energy and on the other meson fields. In
table 1 the soliton energy $E$, its Dirac sea and mesonic
contributions, $E_{\rm vac}$ and  $E_{m}$, as well as the energy of
the valence quark level ($\epsilon_{\rm val}^{R,I}$) are shown for
different values of the constituent quark mass $m$. Also displayed is the
axial charge $g_A$ obtained directly from the axial vector field\cite{Zu94},
$g_A=-(2\pi/g_2)\int dr r^2\left[H(r)+F(r)/3\right]$.
The relative contributions from the Dirac sea and the mesonic part of the
scalar-isoscalar meson field depends on the choice of the
point $r_m$ where the tail for the meson field is fitted. Nevertheless
the total energy and the valence quark energy of the soliton are
stable in a wide range $D/4 < r_m < D/2$. To be definite, we choose
$r_m$ such that $\partial_r \phi(r)|_{r_m} = 0$ which yields
$r_m\approx 0.35 \cdots 0.375 D$.

\begin{table}
\caption{The soliton energy $E$ as well as its Dirac sea and mesonic
contributions $E_{\rm vac}$ and $E_{\rm m}$ for different values of the
constituent quark mass $m$.
Also shown is the energy of the `dived' level $\epsilon_{\rm val}$
and the axial charge $g_A$.}
\vspace{0.3cm}
\centerline{
\begin{tabular}{||l|c|c|c||}
\hline
$m$ (MeV)    ~~~~~~~~~~   &  350 &  400 &  500 \\
\hline
$E$ (MeV)                 & 1125 & 1091 & 1022 \\
\hline
$E^R_{\rm vac}$ (MeV)     & 2271 & 1337 &  883 \\
\hline
$E^I_{\rm vac}$ (MeV)     &  177 &  206 &  223 \\
\hline
$E_{\rm m}$ (MeV)         &-1323 & -451 &  -83 \\
\hline
$\epsilon^R_{\rm val}/m$  &-0.28 &-0.51 &-0.72 \\
\hline
$\epsilon^I_{\rm val}/m$  & 0.15 & 0.15 & 0.12 \\
\hline
$g_A$                     & 0.41 & 0.36 & 0.32 \\
\hline
\end{tabular}}
\end{table}

\vskip0.3cm
\leftline{\it Comparison with the Skyrme model}

Since the energy of the valence quark orbit is negative the baryon
number can well be approximated by the topological current. For that
reason the NJL soliton strongly supports Witten's conjecture that
baryons may be described as solitons within purely mesonic models. We
will now study the explicit connection to the Skyrme model. In a first
step we assume a gradient type expansion for the pseudoscalar field.
Secondly, we adopt the static limit for all other mesons, \ie the
inverse propagators are approximated by the corresponding mass
terms\footnote{In the NJL model the mass of the scalar meson is given
by $2m$.}. This allows one to integrate out these mesons yielding an
extended Skyrme lagrangian\footnote{For notation see ref.~\cite{Po86}}:
\be
 \La &=& \La_2 + \La_4 + \La_6 + \La_{SB}, \nonumber \\
 \La_2 &=& - \frac{f^2_\pi}{4} \tr L_\mu L^\mu, \nonumber \\
 \La_4 &=& \frac{\cos \chi}{32e^2} \tr \left[ L_\mu , L_\nu \right]
	   \left[ L^\mu , L^\nu \right]
           +\frac{\sin \chi}{24e^2} \left( \tr \left( L_\mu L^\mu
	   \right)\right)^2,  \nonumber \\
 \La_6 &=& - \frac{e_6^2}{2} B_\mu B^\mu, \nonumber \\
 \La_{SB} &=& \frac{f_\pi^2 m_\pi^2}{4} \tr \left( U + U^\dagger -2\right).
\label{SkyrmeLagr}
\ee
Here $L_\mu=U^\dagger \partial_\mu U$ and $B_\mu = (1/24 \pi^2)
\epsilon_{\mu\nu\rho\lambda} \tr L^\nu L^\rho L^\lambda$ denote the
left Maurer-Cartan form and the topological current, respectivly. In
the static limit the parameters $e$, $e_6$ and $\chi$ are obtained to be
\be
 \frac{1}{e^2} &=& \sqrt{\left(\frac{\left| 2a-1 \right|}{g_V a}\right)^4 +
       \left( \frac{\sqrt{3}f_\pi}{4 m} \right)^4} \qquad \hbox{with} \qquad
       a=1+\frac{m_\rho^2}{6m^2}, \nonumber \\
 e_6 &=& \frac{N_c g_V}{2 m_\rho},\nonumber \\
 \tan \chi &=& \left(\frac{\sqrt{3}f_\pi}{4 m} \frac{g_V a}{\left|
 2a-1 \right|} \right)^2.
\ee
For a given chiral field $U$ the soliton energy or mass obtained from
the Lagrangian (\ref{SkyrmeLagr}) is given by
\be
 M = \int d^3r \left[ \frac{f_\pi^2}{2}\left( \Theta^{\prime 2}(r) +
 \frac{2\sin^2(\Theta(r))}{r^2} \right) +
 \frac{1}{e^2}\frac{\sin^2(\Theta(r))}{r^2} \left( \Theta^{\prime 2}(r) +
 \frac{\sin^2(\Theta(r))}{2 r^2} \right) \cos \chi \right. \nonumber \\
  \left. - \frac{2}{3e^2}\left( \Theta^{\prime 2}(r) +
 \frac{2\sin^2(\Theta(r))}{r^2} \right)^2 \sin \chi
+ \frac{e_6^2}{8 \pi^4} \Theta^{\prime
 2}(r) \frac{\sin^4(\Theta(r))}{r^4} + m_\pi^2 f_\pi^2
 (1-\cos(\Theta(r)) \right].
\label{Skyrmemass}
\ee

\begin{table}
\caption{Comparison of the mass $M$, eq. (\protect\ref{Skyrmemass}), the masses
of the nucleon ($N$) and $\Delta$-resonance (\protect\ref{nucleonmass}) for
the self-consistent NJL model and the soliton of the extended Skyrme model
(\protect\ref{SkyrmeLagr}). Shown are the results for two different
values of the constituent quark mass $m$. The physical $\rho$ meson
mass is adopted, $m_\rho=m_\rho^{ph}$.}
\vspace{0.3cm}
\centerline{
\begin{tabular}{||l|c|c|c|c||}
\hline
$m$ (MeV)    ~~~~~~~~~~   &
 \multicolumn{2}{c|} {400}   & \multicolumn{2}{c||} {500} \\ \hline
soliton   & eq.~(\ref{SkyrmeLagr}) & NJL & eq.~(\ref{SkyrmeLagr}) &
NJL \\ \hline
$e$ & \multicolumn{2}{c|} {4.65}   & \multicolumn{2}{c||} {5.41} \\ \hline
$e_6$ ($10^{-3}$/MeV) &
 \multicolumn{2}{c|} {12.7}   & \multicolumn{2}{c||} {13.7} \\ \hline
$\chi$  & \multicolumn{2}{c|} {0.221} & \multicolumn{2}{c||} {0.191} \\ \hline
$M$ (MeV)              & 1760 & 3519 & 1714 & 4124 \\ \hline
$M_N$ (MeV)            & 1812 & 3643 & 1770 & 4261 \\ \hline
$M_\Delta$ (MeV)       & 2023 & 4139 & 1995 & 4809 \\ \hline
$M_\Delta - M_N$ (MeV) &  210 &  496 &  225 &  548 \\ \hline
\end{tabular}}
\end{table}

If we perform slow rotations $A(t) \in SU(2)$ on the static soliton
$U(t)= A(t) U A^\dagger(t)$ we can derive the following Hamiltonian in
the space of collective angular degrees of freedom
\be
 H = M + \frac{{\bf J}^2}{2\lambda}
\label{collHamilt}
\ee
with $\lambda$ being the moment of inertia given by
\be
 \lambda &=& \frac{2}{3} \int d^3r \sin^2(\Theta(r)) \left[ f_\pi^2  +
\frac{\cos \chi}{e^2} \left( \Theta^{\prime 2}(r)
+ \frac{\sin^2(\Theta(r))}{r^2} \right) \right. \nonumber \\ &&\quad\left.
- \frac{2\sin \chi}{3e^2} \left( \Theta^{\prime 2}(r) +
\frac{2 \sin^2(\Theta(r))}{r^2} \right)  + \frac{e_6^2}{4 \pi^4} \Theta^{\prime
2}(r) \frac{\sin^2(\Theta(r))}{r^2} \right].
\ee
The mass for the $N$ and $\Delta$ might be extracted from
eq.~(\ref{collHamilt}):
\be
 M_N &=& M + \frac{3}{8\lambda} \nonumber \\
 M_\Delta &=& M + \frac{15}{8\lambda}.
\label{nucleonmass}
\ee
For comparison we have calculated the chiral angle $\Theta(r)$
(\cf fig.~2): on one hand the self-consistent NJL soliton and on the
other the extended Skyrmion obtained from the Euler-Lagrange equation
associated with the Skyrme Lagrangian (\ref{SkyrmeLagr}). We have then
calculated the $N$ and $\Delta$ masses (\ref{nucleonmass}) for the NJL
soliton chiral field as well as for the Skyrmion. As can be seen from fig.~2
using the physical $\rho$ meson mass $m_\rho=m_\rho^{ph}=770$MeV leads to
quite different profiles for the chiral angle $\Theta(r)$.
Accordingly, the soliton mass $M$ and therefore the $N$ and $\Delta$
masses are quite different in the static approximation(\cf tab.~2).
Note that the soliton mass $M$ is significantly larger than the
self-consistent NJL soliton energy (see table 1). This indicates
that the physical $\rho$ meson mass is too low for the validity of the
static limit. Indeed if we increase the $\rho$ meson mass
successively the two chiral angles approach piece by
piece (\cf fig.~2). For $m_\rho = 8 m_\rho^{ph}$ the soliton mass $M$
obtained by substituting the self-consistent chiral angle of the NJL soliton
into the static approximation and the
Skyrmion are in very good agreement as can be seen from tab.~3. Furthermore,
this soliton mass is very close to the corresponding self-consistent
NJL soliton energy which is $E=1162$MeV ($E=1026$MeV) for $m_\rho = 8
m_\rho^{ph}$ and $m=400$MeV ($m=500$MeV), respectively.

\begin{table}
\caption{Same as table 2 for $m_\rho=8m_\rho^{ph}$}
\vspace{0.3cm}
\centerline{
\begin{tabular}{||l|c|c|c|c||}
\hline
$m$ (MeV)    ~~~~~~~~~~   &
 \multicolumn{2}{c|} {400}   & \multicolumn{2}{c||} {500} \\ \hline
soliton   & eq.~(\ref{SkyrmeLagr}) & NJL & eq.~(\ref{SkyrmeLagr}) &
NJL \\ \hline
$e$ & \multicolumn{2}{c|} {5.17}   & \multicolumn{2}{c||} {6.46} \\ \hline
$e_6$ ($10^{-3}$/MeV) &
 \multicolumn{2}{c|} {2.53}   & \multicolumn{2}{c||} {3.15} \\ \hline
$\chi$  & \multicolumn{2}{c|} {0.274} & \multicolumn{2}{c||} {0.274} \\ \hline
$M$ (MeV)              & 1160 & 1162 & 1008 & 1013 \\ \hline
$M_N$ (MeV)            & 1326 & 1328 & 1235 & 1272 \\ \hline
$M_\Delta$ (MeV)       & 1989 & 1991 & 2140 & 2307 \\ \hline
$M_\Delta - M_N$ (MeV) &  663 &  663 & 906  & 1035  \\ \hline
\end{tabular}}
\end{table}

\vskip0.3cm
\leftline{\it Conclusions}

To summarize, we have shown that for an appropriate definition of the
NJL model energy functional the inclusion of all meson
fields in a way compatible with the hedgehog ansatz leads to stable
soliton solutions in a wide range of constituent quark masses.
We would like to put emphasis on the fact that the chiral
radius deviates only mildly from its vacuum expectation value when all
meson fields are included in the evaluation of the self-consistent
soliton. This justifies the frequently adopted approximation $\phi(r)
\equiv 1$.

Including all meson fields with vanishing grand spin
the valence quarks are strongly bound and the baryon number is
completely carried by the polarized Dirac sea. This indicates that
baryons can be described as purely mesonic topological solitons like
the skyrmion. This result has motivated the comparison between the
solitons of the NJL model and a Skyrme type model. The latter has been
related to the NJL model in the limit of large (axial-) vector meson
masses and a gradient expansion for the pseudoscalar field. It has
turned out that these two models possess different soliton solutions
for the physical values of the vector meson mass ($m_\rho^{ph}$).
However, as $m_\rho$ is increased these two solitons become similar in
shape and size. Reasonable agreement is achieved for $m_\rho \ge 8
m_\rho^{ph}$. This indicates that the kinetic term for the (axial-)
vector mesons, here represented by the fermion determinant, carries
important information about the meson fields. We have thus collected
some support for Skyrme type models from a microscopic model for the
quark flavor dynamics. One might argue that purely mesonic models with
explicit (axial-) vector degrees of freedom might be more feasible for
the computation of nucleon properties. However, as one is interested
in the explicit quark structure of baryons it is unremitting to
consider a microscopic model like the one of Nambu and Jona-Lasinio.

\vfill\eject
\setlength{\baselineskip}{12pt}

\newpage

\centerline{\Large \bf Figure captions}

\vskip1cm

\centerline{\large \bf Figure 1}

\noindent
The chiral radius field $\phi(r)$ for $m=400$MeV.
\vskip1cm
\centerline{\large \bf Figure 2}

\noindent
The chiral angle $\Theta(r)$ of the NJL model and the extended
Skyrme model for $m=400$MeV.

\end{document}